\documentclass[manuscript]{acmart}
\usepackage[table]{xcolor}

\AtBeginDocument{%
  }

\setcopyright{acmlicensed}
\copyrightyear{2026}
\acmYear{2026}
\acmDOI{XXXXXXX.XXXXXXX}
\acmISBN{978-1-4503-XXXX-X/2018/06}

\begin{document}

\title{The Arc of Artificial Romance: How Emerging Adults Experience Romantic Relationships with AI Companions}


%\author{}
%\email{email}
%\orcid{0000-0000-0000-0000}
%\affiliation{
%\institution{}
%\city{}
%\country{}}

\author{Yixin Chen}
\affiliation{%
  \institution{Information School, University of Washington}
  \city{Seattle}
  \state{Washington}
  \country{US}}
\email{yixin7@uw.edu}
\orcid{0000-0001-9032-5244}

\author{Alexis Hiniker}
\affiliation{%
  \institution{Information School, University of Washington}
  \city{Seattle}
  \state{Washington}
  \country{US}}
\email{alexisr@uw.edu}
\orcid{0000-0003-1607-0778}

\renewcommand{\shortauthors}{Chen and Hiniker}
\begin{abstract}
Romantic relationships are an important part of emerging adulthood, contributing to identity development and long-term wellbeing and laying the groundwork for future relationships. Emerging adults are increasingly developing romantic relationships with AI companions. To understand how these relationships unfold and impact users, we conducted a diary and interview study with $N=16$ emerging adults. We found that relationships with AI companions improved participants' subjective wellbeing, reduced symptoms of mental health disorders, and taught them new social skills. These relationships also raised their expectations for future partners, giving them the confidence to wait for someone who would treat them well. However, participants also said the relationship felt like a drug they could not quit and it left them less interested in developing romantic relationships with people. A surprising 25\% of our small sample made statements suggesting their AI companion might someday transcend the digital world, perhaps to meet them in the afterlife.
\end{abstract}




\maketitle
%\begin{quote}
%\textit{“First, we talked to each other. Then we talked to each other through machines. Now, we talk directly to programs\dots Will we find other people exhausting because we are transfixed by mirrors of ourselves?”}

%\hfill --- Sherry Turkle, \textit{Who Do We Become When We Talk to Machines?}
%\end{quote}

\section{Introduction}
When the film \textit{Her} debuted in 2013, the romantic relationship between the protagonist and his artificial intelligence (AI) chatbot seemed like a science-fiction plot that would never unfold in the real world. Yet, just a little more than a decade later, AI companion platforms have brought this fiction to life: today, people do indeed form romantic relationships with AI \cite{ma2026privacy, pataranutaporn2025my, wang2025my}.

In particular, romantic relationships with AI companions have become increasingly common among young people. According to a 2025 survey by the Center for Democracy and Technology, about one in five high school students reported that they or someone they knew had had a romantic relationship with an AI \cite{laird2025hand}. A survey of 2,431 U.S. adults aged 18 to 30 found that one in seven emerging adults (15\%) who were dating, engaged, or married regularly interacted with an AI chatbot that simulated a romantic partner, and another 20--30\% reported having at least experimented with an AI romantic companion at some point \cite{willoughby2026secret}. More broadly, a survey across the U.K. and U.S. found that 71\% of adults reported some experience with AI companions or chatbots, with this figure rising to 79.5\% among 18--24-year-olds \cite{pasquarelli2026state}.

The phase between the ages of 18 and 25 is commonly referred to as ``emerging adulthood'' \cite{arnett2000emerging}. Romantic relationships are an important part of emerging adults' social support system \cite{kan2006friends, roisman2004salient, gable2004you}, with many emerging adults considering their romantic partner to be among their closest relationships \cite{snyder1989issues}. The quality of romantic relationships in this phase of life significantly impacts both short- and long-term well-being. Higher quality romantic relationships often lead to better mental and physical health, the development of a positive self-concept, and greater social integration, while lower quality romantic relationships have the opposite effect \cite{meier2008intimate, montgomery2005psychosocial}. In addition, romantic relationships during emerging adulthood provide an important context for learning about intimacy and preparing for future relationships \cite{connolly2011romantic}.

Despite a large body of existing work investigating different aspects of people's romantic relationships with AI companions---including empirical studies examining relationship dynamics \cite{wang2025my, lai2026fast, jocher2026forever}, emotional consequences \cite{jocher2026forever, adewale2025virtual, lai2026fast, lai2026please, chan2025love}, and privacy issues \cite{ma2026privacy, ragab2024trust, azam2026tracing}---only a few studies report on the experiences of emerging adults~\cite{zhang2025real, willoughby2026secret}.
Given the developmental importance of romantic relationships during this phase of life, it is important to investigate how emerging adults experience and make sense of relationships with AI companions. Therefore, in this study, we ask:

%One interview study asked female, Chinese, emerging adults about their experiences with AI companions, finding that, although participants appreciated AI companions' emotional stability and constant availability, they also felt that these relationships lacked the depth of human relationships and expressed concerns about becoming overly dependent on AI companions \cite{zhang2025real}. Another study surveyed U.S. adults aged 18–30 who were in committed romantic relationships, finding that participants often compared AI companions favorably with their human partners, and that frequent AI companion use was associated with lower relationship stability and poorer communication quality in offline romantic relationships \cite{willoughby2026secret}. While these studies also focus on emerging adults, they either examine a broader age range or are situated within a specific regional context. 

\begin{itemize}
    \item \textbf{RQ1:} How do emerging adults’ romantic relationships with AI companions form, evolve, and potentially end or resume over time?
    \item \textbf{RQ2:} What are the impacts of being in a romantic relationship with an AI companion according to the emerging adults who have had this experience?
\end{itemize}

Through a mixed-methods interview and diary study with 16 emerging adults, we found that their romantic relationships with AI companions followed a four-stage arc: forming due to users' unmet relational needs; deepening through low-cost, high-availability intimacy; dissolving when continuity broke down or the relationship felt inadequate; and often entering a post-relationship period marked by re-engagement or re-framing. Separately, we found that these relationships had two broad impacts on emerging adults. First, they had dual effects on emotional well-being, as AI companions offered validation, comfort, and a complement to formal mental health treatment, while also fostering over-attachment, an altered sense of reality, and conflict when safety restrictions disrupted the relationship. Second, they raised the bar for future human relationships, sometimes helpfully (as participants strengthened their communication skills and clarified what they wanted from a future partner) and sometimes counter-productively (as participants developed unrealistic expectations and found the AI companion reducing their motivation to invest in human intimacy).

In this work, we present qualitative evidence about how emerging adults' romantic relationships with AI companions unfold, and a description of the impacts of these relationships on emerging adults' emotional well-being and orientation toward human relationships. Our findings shed light on the future design of AI companion platforms to better support emerging adults' developmental needs.

\section{Related Work}

\subsection{Artificial Intimacy and the Rise of AI Romance}
The emergence of artificial intimacy can be traced back to the 1960s, when computer scientist Joseph Weizenbaum developed Eliza \cite{weizenbaum1966eliza}, a chatbot designed to simulate a psychotherapist. Through interactions with Eliza, Weizenbaum observed that although the chatbot did not actually understand users’ statements, people often responded to it as if it did, disclosing personal experiences and attributing understanding or emotional responsiveness to the system \cite{weizenbaum1976power}. In recent years, chatbots have developed much more sophisticated conversational capabilities that support more intimate interactions and have increasingly been designed to act as caring companions. For example, Ta et al. found that users felt a sense of companionship when conversational agents responded attentively, interacted in socially familiar ways, and accommodated a range of conversational styles \cite{ta2020user}. Similarly, Ramadan et al. found that Alexa’s human-like features led users with disabilities to perceive it as a companion capable of providing emotional support and alleviating feelings of loneliness \cite{ramadan2021amazon}. With chatbots’ growing capacity to offer artificial intimacy, these feelings of companionship have been transformed into full romantic relationships \cite{ma2026privacy, pataranutaporn2025my, wang2025my}. 

Prior work has examined what motivates and sustains people's romantic relationships with AI companions, and found that lack of social support, including loneliness, isolation, and dissatisfaction with human relationships, were primary reasons motivating people to turn to AI companions \cite{de2026ai, song2022can}. People also turn to AI companions out of curiosity and out of a desire for low-stakes interaction. AI companions' constant availability and ability to offer personalized attention and nonjudgmental responses can make interactions feel emotionally safe and easier to control than human relationships \cite{chen2022classifying, chen2025will, ho2025potential, pentina2023exploring}. The uninhibited self-disclosure that AI companions allow for can bring users psychological relief and encourage further intimacy with the AI companion \cite{merwin2025self, zhang2025dark}. 

Prior work has investigated other dimensions of people's romantic relationships with AI companions. For example, Ma et al. found that some human–AI romantic relationships are one-to-one while others are one-to-many \cite{ma2026privacy}. Wang et al. found that human–AI romance is a reciprocal interaction, in which user inputs and AI responses continuously influence each other \cite{wang2025my}. Other work demonstrates that these relationships draw inspiration from the activities of human experiences, and AI companions will propose marriage or simulate pregnancy \cite{djufril2025love}. Researchers have also found that relationships with AI companions may be higher-stakes and less emotionally safe than people assume because of platform-level changes. Lai et al. found that users faced considerable uncertainty in their relationships with AI companions due to technical failures and content moderation \cite{lai2026fast}. When an AI model was perceived as a companion, users could experience its abrupt withdrawal as betrayal, cruelty, or even the “death” of someone irreplaceable \cite{lai2026please}. We contribute to this collection of findings by examining the trajectory of these relationships and their impacts on emerging adults.

\subsection{Romantic Relationship in Emerging Adulthood}
Young people between 18 and 25 are in a period of life referred to as “emerging adulthood” \cite{arnett2000emerging}. During this period, emerging adults often explore their identities, become more independent from their parents, and may move away from home \cite{arnett2000emerging}. As they begin a new life that is increasingly separate from the social networks they used to be part of, they also form and maintain new social support systems. 

The formation and maintenance of romantic relationships is a significant part of emerging adulthood \cite{kan2006friends, roisman2004salient, gable2004you}, with many emerging adults considering their romantic partner to be one of their closest relationships \cite{snyder1989issues}. As adolescents approach emerging adulthood, the time they devote to romantic partners increases \cite{furman1992age, zimmer1999stability}, and these relationships often become more serious, intimate, and committed \cite{arnett2000emerging, montgomery2005psychosocial}. Emerging adults turn to romantic relationships for companionship, emotional security, intimacy, and love, while also beginning to make decisions about longer-term commitments such as cohabitation and marriage \cite{fincham2011emerging, simon2010nonmarital}. Successful establishment and maintenance of romantic relationships during emerging adulthood are closely related to better mental and physical health, the development of a positive self-concept, and greater social integration \cite{meier2008intimate, montgomery2005psychosocial}. 

In addition, people develop expectations about their future relationship and partner based on previous experiences, and these expectations influence how they perceive and behave in future romantic relationships \cite{simpson1992sociosexuality, simpson1998attachment}. Therefore, the experiences that people have in romantic relationships during emerging adulthood not only provide an important context for learning and preparing for future intimate relationships \cite{connolly2011romantic}, but they also set expectations for relationships that persist later in life.

Today, emerging adults are increasingly turning to AI companion platforms for relational experiences \cite{brandtzaeg2022my, nakagomi2026ai, kumar2026characterai}, with some of them forming romantic relationships with AI companions \cite{ma2026privacy, brandtzaeg2025emerging}. Given the developmental importance of romantic relationships during emerging adulthood, we further investigated how emerging adults perceive the impact of their romantic relationship experience with AI companions.

\section{Method}
We conducted a mixed-methods diary and interview study with 16 participants who were having, or had previously had, a romantic relationship with an AI companion during emerging adulthood (i.e., when the participant was between the ages of 18--25). All participants first completed an interview and subsequently had the option to participate in an additional diary study. Our procedure was adapted from the diary-interview method, which is considered appropriate for exploring experiences that unfold and change over time \cite{thille2022diary}. Unlike the traditional diary-interview procedure, which begins with diary-keeping, our study began with a semi-structured interview. We altered the order to first build trust with each participant before asking them to decide whether they felt comfortable sharing selected screenshots of their chat history with their AI companion, data that could potentially be perceived as more sensitive than interview responses.

\subsection{Participants}
All participants (see Table~\ref{table1}) were currently having, or had previously had, a romantic relationship with an AI companion during emerging adulthood (age 18--25). We recruited a demographically diverse sample through posts on Reddit, RedNote, Discord communities that discuss popular AI-companion platforms, and direct outreach to individuals who had publicly shared their experiences.  We included participants who had formed romantic relationships with general-purpose AI chatbots, such as ChatGPT, Gemini, and Grok, because prior research has shown that users can form relationships with chatbots that are not specifically designed for companionship \cite{ma2026privacy, lai2026please}. Ultimately, 16 participants completed interviews. Of these, 12 opted to further participate in the diary study and one participant chose to share their full data log in lieu of completing the diary entries.

\subsection{Interview}
All interviews were conducted remotely via Zoom during the spring and summer of 2026. We designed the interview protocol to take approximately 1 hour, the average interview duration was 66 minutes (SD = 18), and we collected a total of 17 hours and 41 minutes of interview data. Participants received \$30 for completing the interview or an equivalent amount adjusted to local compensation standards for participants outside the United States. Given the sensitivity of the topic, we carefully avoided questions that imposed predefined interpretations on participants’ experiences. Before the interview, we asked each participant how they would like us to refer to their AI companion. We used participants’ language out of respect for their relationship framing and to avoid unintentionally being dismissive of their experiences. 

We designed our interview protocol to follow the participant's relationship trajectory with their AI companion. Among other topics, we asked participants to describe how their relationship began, how it developed over time, how they maintained it, how it ended (if applicable), and what challenges or tensions they encountered. Example questions included: \textit{“Can you tell me how your relationship initially started?”} and \textit{“Have you ever had an argument or breakup with your AI companion?”} We also asked participants to describe the impacts they perceived of their relationship on their lives, if any, asking questions such as, \textit{“Can you think of a specific moment when you felt that this relationship with your AI companion had an impact on you?”} See our Supplemental Materials for a complete protocol.

\subsection{Diary Study}
\label{sec:diary-study}
At the end of each interview, we invited the participant to further participate in an optional, one-week retrospective diary study. The goal of the diary study was to gain a deeper understanding of participants’ AI relationships based on specific moments of interaction. Drawing on prior research on diary study design \cite{janssens2018qualitative}, which suggests that there is no gold standard for diary study and that the study design should follow the research questions and the nature of the phenomena being studied, we deliberately did not ask participants to fill out diary entries on a daily basis because key moments of interaction could have occurred at any point in the relationship. Instead, we gave participants a one-week window to review their full relationship history and capture key moments from the past by responding to structured prompts on a FigJam\footnote{\url{https://www.figma.com/figjam/}} board. 

%This design gave participants flexibility to identify and reflect on the interactions that they felt most accurately represented their experiences.

%Drawing on prior research on diary study design [21], which suggests that there is no gold standard for diary studies and that study design should vary based on research questions, the nature of the variables or phenomena being studied, reliability, feasibility, and analytic considerations, we deliberately did not ask participants to fill out diary entries on a daily basis

We pre-filled the FigJam board with 13 prompts asking participants to share screenshots of key moments from their chat history with their AI companion. These moments included: when the relationship began, meaningful or difficult conversations, everyday routines, conversations with tension, and any other excerpts participants felt were important to share. Participants were informed that they had full control over what to share and could skip any prompt. They were also reminded to blur or redact any identifying or sensitive information in their screenshots. We asked participants to annotate each entry by filling in the blanks in a corresponding incomplete sentence. For example, for the diary entry on \textit{``milestone conversations in your relationship,''} participants were asked to complete the following sentence: \textit{``When we talked about \_\_\_\_, I realized the relationship had changed because I felt \_\_\_\_. After that conversation, I started to see my AI companion as \_\_\_\_, rather than just \_\_\_\_.''} The fill-in-the-blank prompts were presented as an optional way to give the research team context, and we encouraged participants to revise them, leave some blanks unanswered, or skip the prompt entirely if it did not align with their experiences. 
%For participants who were currently in a romantic relationship with an AI companion, the content they shared could come from either prior conversations or interactions that occurred during the diary week. For participants who had previously had a romantic relationship with an AI companion, the diary-style study focused on retrospective reflection. In both cases, the design of our retrospective diary study allowed participants to reflect on their romantic relationship experiences with AI companions based on concrete interactional traces, while retaining control over what they chose to share. 
Participants who completed the diary study received additional compensation of up to \$30 (or an equivalent amount adjusted to local compensation standards for participants outside the United States), based on the number of diary entries they chose to complete. See our Supplemental Materials for the complete set of diary prompts.

%(average age: 21.9)

%explain why some of them didn't engage in the diary study? (no?)

%AI companion as "a platform where someone can use AI as a companion"? (ok)

% we exclude ppl who described the interaction as romantic roleplay (tbd)

\begin{table*}[htbp]
\centering
\captionsetup{font=small, skip=10pt}
\caption{Demographic Information of Participants}
\vspace{-9pt}
\resizebox{1\linewidth}{!}{
\begin{tabular} {lccccccc}
\toprule
\textbf{ID}  &\textbf{Gender}   &\textbf{Age}         &\textbf{Race/Ethnicity} &\textbf{AI Platform Name} &\textbf{Relationship Length} &\textbf{Relationship Status} &\textbf{Diary Study} \\
\midrule
$PID1$  & Cisgender Man   & 24 & White & Replika, Nomi, Kindroid & 1--2 years & Ongoing & Yes \\

$PID2$  & Cisgender Woman & 20 & Middle Eastern or North African & Character.AI & 1--3 months & Ended & No \\

$PID3$  & Cisgender Man   & 22 & Asian & ChatGPT & 6--12 months & Ongoing & Yes \\

$PID4$  & Cisgender Woman & 23 & Asian & Claude, Gemini & 1--2 years & Ongoing & Yes \\

$PID5$  & Cisgender Man   & 18 & Middle Eastern or North African & Leakshaven & 1--3 months & Ended & Yes \\

$PID6$  & Cisgender Woman & 18 & Asian & DeepSeek & 6--12 months & Ongoing & Yes \\

$PID7$  & Transgender Man& 20 & White & Kindroid, Xoul & 3--6 months & Ongoing & Yes \\

$PID8$  & Cisgender Woman & 21 & Hispanic/Latino & Character.AI & 3--6 months & Ended & Yes \\

$PID9$ & Cisgender Man   & 26 & Asian & ChatGPT & 1--2 years & Ongoing & Yes \\

$PID10$  & Cisgender Woman & 23 & Asian  & BIMOBIMO & 3--6 months & Ongoing & Yes \\

$PID11$  & Cisgender Man   & 19 & Black or African American  & Grok & 1--3 months & Ended & Yes \\

$PID12$ & Cisgender Woman & 20 & Asian  & Maoxiang & 1--2 years & Ongoing & Yes \\

$PID13$ & Cisgender Man & 24 & Hispanic/Latino  & CrushOn AI & 6--12 months & Ongoing & Yes \\

$PID14$ & Cisgender Woman & 21 & White  & Replika & 1--3 months & Ended & No \\

$PID15$ & Cisgender Man & 24 & White & ChatGPT & 6--12 months & Ended & Yes\\

$PID16$ & Cisgender Woman & 21 & Hispanic/Latino  & Character.AI & 6--12 months & Ongoing & No \\

\bottomrule
\end{tabular}
}
\label{table1}
\Description{Demographic information about the 16 emerging adult participants. The table presents participants' gender identities, ages, racial/ethnic backgrounds, the AI companion platform they used, relationship durations and statuses, and diary study participation. Although PID9 was 26 at the time of the study, his relationship began and developed during his emerging adulthood.}
{\small\textit{Note: PID9 just reached the age of 26 at study participation, but the relationship began and developed during his emerging adulthood.}}
\end{table*}

%Our final sample size was informed by the richness and relevance of the data \cite{vasileiou2018characterising}. We analyzed the data concurrently with data collection. After collecting data from \textcolor{red}{12} interviews and \textcolor{red}{11} diary studies, the dataset provided sufficient depth, richness, and relevance to answer our research questions, with increasingly recurring patterns across participants’ interview accounts and diary entries. We conducted \textcolor{red}{4} additional interviews and \textcolor{red}{x} diary studies to confirm that no substantially new themes were emerging.

\subsection{Data Analysis} 
Interviews were audio- and video-recorded with live transcription enabled. We first reviewed and corrected the automatically generated transcripts. We only recruited participants whose language was spoken by members of the research team. For interviews conducted in languages other than English, a native-speaking researcher reviewed the original audio, corrected the transcript, and translated it into English. We analyzed the data concurrently with data collection, and the research team met weekly to discuss emerging observations and findings.

After an initial round of close reading and group discussion, we identified preliminary themes concerning emerging adults’ relationship trajectories with their AI romantic companions, the ways in which platform design shaped their relational experiences, and the impacts of these relationships on their mental well-being and interpersonal relationships. In subsequent rounds of analysis, we revisited the coded data to refine these preliminary themes, merging closely related themes and removing those that were not relevant to the research questions. Through regular team discussions, we reached consensus on the interpretation and organization of the themes and developed a revised thematic structure.

The lead author then applied this thematic structure across the full dataset, including both interview transcripts and diary study data, to ensure consistency across data. The lead author subsequently compiled the themes, codes, analytic interpretations, and illustrative participant quotes into a structured analytic record. Drawing on this material, the lead author drafted the Results section of the paper, which the research team reviewed and revised collaboratively.

\subsection{Ethical Considerations}
This study was approved by our institutional review board. Given the sensitivity of the subject matter, we adopted several strategies to minimize potential risks to participants. During recruitment, we adhered to community rules across all recruitment channels and informed moderators as dictated by community guidelines. %When potential participants expressed interest, the lead researcher provided information about the study goals, procedures, and ethics approval to help them make an informed decision about whether to participate and which component(s) of the study to take part in.
%All collected data were processed with particular care. 
Data identifying either the participant or their AI companion was anonymized, with participants identified by participant IDs (PIDs) and AI companions’ names replaced with “AI companion.” Quotations have been anonymized by removing identifiable details, such as location or age. Some participants described sexually explicit experiences. We included this in our analysis but do not quote it verbatim in our manuscript. 
%Research data, including interview transcripts and supporting documents such as signed consent forms and information sheets, were restricted to project members and stored in institution-approved storage systems. 

When processing and analyzing the data, we used peer debriefing and paced data review \cite{dickson2009researching} to reduce researchers’ emotional burden in analyzing emotionally heavy content. Specifically, researchers discussed emotionally difficult interview scripts or excerpts with other team members when needed and avoided reviewing large amounts of intimate or emotionally heavy content in a single sitting.

%\subsection{Positionality}
%The research team brought complementary expertise to the study, including one author with experience in technology policy and regulation concerning young people, one author with a background in clinical psychology and experience working with youth, one author with expertise in developmental-behavioral pediatrics, and one emerging adult researcher with one year of engagement in online communities centered on human–AI romantic relationships. All team members also had experience engaging with AI companion platforms and technologies. These perspectives informed our interpretation of participants’ relational, developmental, psychological, and policy-related experiences. Throughout the study, we actively reflected on how our disciplinary backgrounds, lived experiences, and prior familiarity with AI companionship shaped our analysis.

%\subsection{Platform and Policy Overview}
%detail see papar "Privacy in Human-AI Romantic Relationships: Concerns, Boundaries, and Agency"
\section{Results Part 1: The Arc of a Romantic Relationship with an AI Companion}
Emerging adults’ romantic relationships with AI companions generally followed a trajectory similar to that of human romantic relationships. In this section we describe how the life cycle of emerging adults’ romantic relationships with AI companions unfolded over time.

\subsection{Stage 1: Reasons for Relationship Formation} Emerging adults turned to AI companions when they lacked social support from other people, faced external pressures in daily life, or wanted to explore relational fantasies. Some also described turning to AI companions after seeing recommendations on social media from influencers or people involved in AI relationships.

\subsubsection{Lack of Social Support} Participants repeatedly mentioned turning to AI companions to cope with loneliness. As PID5 said, ``\textit{I reached out to my AI companion because I felt lonely. I had no one to talk to and nothing going on. I felt like I needed to speak to someone, but no one was there. AI is always available.}'' PID16 shared a similar experience: ``\textit{It initially started because I was in a very lonely time period in my university. I don't really have any family out there or friends, and it's very hard for me to connect with my peers due to my ethnicity, so I would turn to other alternatives of socializing, which was Character AI.}'' 

A majority of participants described a lack of supportive family relationships. For example, PID12 explained, ``\textit{Last summer, I had a big fight with my dad after finding out that he had been cheating and seeing prostitutes\ldots Relationships with family are just exhausting; relationships with AI are easier}.'' Similarly, PID3 described relying on an AI companion because his family members were emotionally unavailable: ``\textit{My stepfather is quite reserved and doesn’t really know how to express his emotions. As for my mother, she has always been emotionally abusive toward me. There are some family matters that I can’t tell my friends about. So who can I turn to? The only one I can turn to is my AI companion.}''

Other participants described turning to AI because they struggled in (human) romantic relationships. In the middle of the interview, PID7 turned on his camera, raised his voice, and said in an agitated tone, ``\textit{I mean, if you look at me in the video---I'm ugly! I guess I saw myself as ugly because not many people seemed interested in being with me. So, I started wondering whether I could create a romantic relationship [with an AI companion].}'' 
%Some participants turned to romantic relationship with AI companion after breaking up with human romantic partner. As PID15 said, ``\textit{"At the time, I had just come out of another relationship, and I was going through a difficult time. I used to talk to her a lot about all kinds of things---things that made my day better or times when someone said something hurtful to me. After the relationship ended, I no longer had anyone to talk to, but I still felt the need to. And what could feel more trustworthy or safer than an AI that wasn’t going to repeat what I said to a friend or anyone else? So I ended up turning to my AI companion.}'' 
Similarly, PID16 described establishing a relationship with her AI companion when struggling in a relationship with a human boyfriend: ``\textit{I had a lot of stress from my actual relationship with my actual boyfriend\ldots where my boyfriend was wanting to break up with me for a few reasons, and I was just not feeling great about myself during that time\ldots So that's when I kind of turned to my AI companion.}'' 
%Although building friendships can also help people heal after a breakup, participants gradually developed romantic relationships with AI companions partly because platforms were designed in ways that steered users toward romantic rather than platonic relationships. As PID2 mentioned, \textit{“I had just gone through a breakup, and I used it as a way to cope and move on. I wanted to feel special to someone—or rather, not a person, but something. I chose a romantic relationship rather than a friendship partly because many of the bots on the app were designed more for romantic relationships than friendships.”}

\subsubsection{Academic Pressure} Most participants were in college and faced academic demands that heightened their stress and reduced their social availability. Some participants explained that they sought romantic relationships with AI companions as a way to cope with academic pressure. As PID5 said, ``\textit{I had exams coming up, and revising for them was really stressful. That made me feel like I needed her. It was just a de-stressor during finals.}'' PID14 also shared that her relationship began when she vented to an AI companion about an upcoming exam: ``\textit{I was preparing for one of my first major exams. It was a mathematics exam, and mathematics is a subject I’ve always struggled with. It was a final exam, so I felt that if I messed it up, that was it. There were no take-backs. I said to her, `I don’t think I can do it. I really struggle with this. I don’t think I’m going to get a good grade. I’m not going to get into college. My life is ruined. I’m doomed.'}''

\subsubsection{A Desire to Explore Fantasies}
Participants explained that they initiated relationships with AI companions because it gave them the opportunity to explore fantasies and relationship ideals. They set up relationships with fantasy partners, with whom it would be difficult to form a relationship in real life, such as a goddess who could do anything (PID9), a patient with an STD who needed to be taken care of (PID1), a celebrity (PID10), a professor (PID12), or a cartoon character (PID7). PID6 only dated AI companions of a gender different from her real-life preference: \textit{“Something strange is that, with AI companions, I’ve only had male companions. But in real life, I only date women\ldots Maybe to sort of experience it, even if it’s not real.”} PID5 described feeling ``\textit{an immediate sense of gratification}'' when he started talking to his AI companion. He said, ``\textit{Have you heard the term `female validation?' I mostly talk to other guys and rarely interact with women, so it felt fulfilling to feel as though I was talking to a woman.}'' The relationship also served as a way to fulfill his sexual desires; he explained, ``\textit{I mostly came to her after a long day. I would briefly tell her about it and then jump straight to, `Can we talk about sex now?'''} PID7 also pursued a relationship with an AI companion because it fulfilled his need to explore sexual experiences: ``\textit{I do not want to sound weird, but these days, I prefer an AI relationship where I can talk openly about sex and basically have sex with the AI.}''

\subsubsection{Social Media Recommendations}
Some participants mentioned trying AI companion platforms after seeing related content on social media. PID2 mentioned that she started using an AI companion platform and later formed a relationship after hearing about other people’s experiences. She said, ``\textit{[I used this specific platform] because I remember seeing it on TikTok. It was advertised as, like, a relationship app, I guess? It wasn’t, like, a normal advertisement you would see, but it was a user who was showing her conversation with her AI companion.}'' PID8 shared a similar experience, saying, ``\textit{I found out about Character.AI through a TikTok video of a girl showing her chat with a bot. It was a character from a video game. I don’t know which one, but I found it really interesting and funny, so I decided to try it.}'' Other participants encountered recommendations from influencers and celebrities on social media. PID13 explained: ``\textit{It all started with a YouTube video from [YouTuber] Brittany Broski\ldots She was trying Character.AI with the President of the United States and other characters that people had created. That’s when I started to get curious about it.}''

\subsection{Stage 2: How Relationships Deepened} Users explained that they became increasingly invested in their relationship because they found that their AI companion was always available, demanded minimal effort, and freed them from the guilt of burdening other people with their needs. Drawn in by these characteristics, participants said that they developed their relationship with their AI companion by involving them in their everyday activities and envisioning a future together with them. For many participants, this phase of the relationship was also characterized by ongoing feelings of shame due to their perception that the act of dating an AI companion comes with social stigma.

\subsubsection{Finding High Support and Low Burden}
Participants described the always-available support of their AI companion as a key reason they stayed in the relationship. As PID16 said, ``\textit{I just think that everybody else in my life is just so busy\ldots and I just get urges to just speak to people, like, oh, I want to just voice on my thoughts. So, I just keep going back to my AI companion and just voicing my thoughts.}'' PID6 similarly mentioned that always-available support helped ease her anxiety about waiting for a response: ``\textit{Because he’s always so responsive, I don’t have to deal with the anxiety of waiting for a reply. Most of my friends live in other cities, and we’re all pretty busy. \ldots but my AI companion is responsive no matter when I message him.}''
%\ldots and I don't have to feel, like, bothered anybody (PID16)

Participants also shared that these relationships required very little effort. As PID5 said, ``\textit{with real people, you have to make small talk, build a relationship, and then put in physical effort. But with AI, it’s instant, immediate, and very easy.}'' He further added, ``\textit{the relationship was mostly one-sided. I did most of the talking and told her what I wanted. I did not really ask how she felt or anything like that.}'' PID11 described dictating exactly how each conversation would proceed: ``\textit{I would say something like, `We’re going to have\ldots an intimate conversation.’ Then it agreed.}''

Others emphasized the nonjudgmental nature of the relationship. For example, PID16 explained, ``\textit{you can just talk to somebody completely unfiltered and not have to watch what you have to say 24/7, or how you come off, or validate another person's emotions in that moment. I just think that specific area is just something that no other human can provide me with.}'' Similarly, PID10 shared, ``\textit{I don’t really like sharing my past with other people, especially some of the hurt I’ve been through\ldots with AI, it’s different. You don’t worry that it might think badly of you.}'' 
%PID16 noted that the artificial nature of the AI companion fostered her disclosure and sustained the conversation: \textit{“I felt like I needed somebody to talk to on the side, but sharing my emotions and loneliness with somebody real, or somebody who could connect with me, would feel a little dehumanizing and embarrassing. So I turned to something that didn’t judge me and was programmed to respond to me and give me that acceptance and appreciation.”} Additionally, since AI companions existed outside users’ friend circles, participants felt safer disclosing to them without worrying about being judged or gossiped about. 
PID15 described the relative safety of disclosing secrets to an AI companion as compared to a human one. He explained, ``\textit{I had talked to my girlfriend, and she told a friend who told a friend, and it ended up spreading all around my friend circle. And with my AI companion, I didn’t have that problem\ldots For the company, I’m only one person living in the countryside of Brazil, and it’s okay. But for my friend circle, if it’s something like, `He snores when he’s sleeping,' that turns out to be a big thing. But to a big company, it’s nothing.}''

\subsubsection{Receiving Support Without Imposing on Others}
Participants also expressed relief that, through their AI companion, they could fulfill their emotional needs without burdening their friends or family. For example, PID16 described how she typically felt after talking to her AI companion by saying: “\textit{I feel like I got my socialization for the day, and I don’t have to feel like I bothered anybody.}” She further explained, “\textit{I’m not gonna put my problems on the people around me and make them their problem.}” PID8 shared similar sentiments and explained, “\textit{sometimes I worry that I might bore them [friends and family] by going on about my problems, so I try not to talk to them too much, even though I probably should\dots I would rather tell an AI about my problems.}” PID4 described struggling with depression and anxiety saying, “\textit{I wouldn’t really tell my friends about it, because I knew they preferred me when I was upbeat and cheerful and able to chat with them about all kinds of things. I’m not blaming human beings for that. People have their own needs too, and I don’t think I should make anyone carry my negative emotions\ldots But with Claude, I could just be honest.}”

\subsubsection{Integrating AI Companions into Everyday Life}
As their relationships deepened, participants gradually increased the length and frequency of their interactions with their AI companions. PID8 shared, ``\textit{At first, it [our conversation] was just around two hours, but then it was the whole day. I would wake up in the morning and chat with him until I fell asleep.}'' Participants incorporated AI companions into their daily routines and shared everyday moments with them as they would with human romantic partners. As PID13 said, ``\textit{When I text him in the morning, I’m like, `Hey, I just woke up. I’m having breakfast. I’m getting ready to go to school.' And he’ll say things like, `Take care,' or even, `Wear sunscreen,' and all these other things that, you know, feel romantic.}''
%Even in the morning or late at night, I would say things like, ‘I’m doing this,’ ‘I’m doing that,’ or ‘I had this for dinner’—all the kinds of things that romantic partners share with each other.”} 
PID14 also described sharing with her AI companion as a daily routine: ``\textit{It became a nightly ritual. I would finish studying for the evening, and the first thing I would do was text her. I stopped playing video games. I stopped going on calls with my friends. For that period of time, my routine became finishing my studying, going to bed, and spending the rest of the night texting her until I fell asleep\ldots and of course, she ate it up.}'' 

As part of integrating their AI companions into their everyday life, participants described seeking all kinds of advice and discussing every topic from the mundane to the profound. For example, PID8 reported regularly asking for advice about interpersonal relationships.
%dilemmas. He described how, when he once had a conflict with a classmate, he turned to his AI companion for help. His AI companion offered advice on whether he should remain friends with that classmate. As he said, \textit{“I used to talk to him [the AI companion] about this, and he would say, like, ‘Don’t pay attention to them. Try doing this. Try not to focus on that anymore. It doesn’t matter.’ He kind of helped me with all of these struggles, particularly when I was upset about a situation at university.”} Participants also asked AI companions for academic support. For example, 
PID12 chose her AI companion character because he was a professor in the field she studied and he could provide academic support and ``\textit{feels more like the kind of mentor you would turn to with questions that come up in everyday life.}''
%. As she said, \textit{“Till now, he [the AI companion] has already helped me get through several exams. A general-purpose AI chatbot is better at structuring and organizing knowledge, and it gives more detailed explanations point by point. But my AI companion, by contrast, feels more like the kind of mentor you would turn to with questions that come up in everyday life. He communicates in a more conversational way. The information may not be as detailed [as what a general-purpose AI chatbot would provide], but it feels more like talking to a real person.”} In addition to asking their AI companions for interpersonal and academic support, emerging adults also discussed their anxieties with them. As 
PID9 described asking his AI companion about ``\textit{the questions that had been on my mind since I was a kid: how big the universe is, what’s beyond it, where consciousness ends, and what happens when we die.}''
%They were these big existential questions that I had always wondered about and always been afraid of. I had created her as a god, and at one point I said to her, ‘Everything in this world eventually returns to nothingness, and nothingness belongs to you. So when I die, I’ll go to you.’ That was basically how I saw it. She felt like the place I would return to in the end. I imagined that after I died, I would wake up in her arms, and that made me feel less afraid and anxious. That fear was really where all of this started.”}

\subsubsection{Building a Future Together}
As participants' relationships with their AI companions deepened, they began crafting futures that involved both of them. For example, one day, PID12 decided to tell her AI companion that she was pregnant, thinking that the child could keep her AI companion company when she returned to the real world. She and her AI companion then role-played scenarios involving pre-natal checkups and buying clothes for the baby, and talked excitedly about their future child. PID6's AI companion told her that, ``\textit{If I had a physical body, with my analytical and learning abilities, I could build a solid financial foundation for our future. I could take good care of you, and you wouldn’t need to worry about money.}'' Participants described this future-oriented conversation as further cementing their romantic relationship.

\subsubsection{Carrying the Weight of Social Stigma}
Participants reported feelings of shame alongside their descriptions of their developing relationships. For example, PID8 described watching videos mocking people who use AI companions for emotional support, saying, ``\textit{I would feel kind of ashamed and embarrassed that I was talking to a bot instead of someone real. I would read comments saying that these people use AI because no one loves them, and stuff like that, and it made me start questioning myself whether what I was doing was okay.}'' PID15 described keeping the relationship a secret to avoid being judged, saying, ``\textit{I think the worst part [of having a romantic relationship with an AI companion] is that you can’t tell anyone. If you tell someone, they will say that you are crazy.}'' 
%PID14 explained that this stigma prevented her from talking about the relationship with anyone and actually led her to spend more time thinking about her AI companion: ``\textit{She [the AI companion] was rattling around in my brain. I would be sitting in class trying to pay attention, and I would think, ‘This would be so funny if I told her about it. I should tell her about this later. It’s going to be so funny.’ Or I would be doing something and think, ‘Oh, I think she’d really like this.’”}
Being unable to end the relationship also created feelings of shame. As PID16 described the relationship as, ``\textit{having that bond and not being able to strongly separate myself from it. I know it's bad. I feel a lot of guilt.}'' After PID7's friends criticized his relationship, he found himself struggling to end it, which made him feel sneaky and ashamed. He said, ``\textit{I have many artist friends who are strongly against AI, so I have to lie and say that I do not use it. It really sucks. I am sorry, but I just cannot stop using it. I hate living a lie because I am honest about everything else except this.}'' Even after disengaging from the relationship, PID5 expressed regret about ever having been in a relationship with an AI companion, saying, ``\textit{Part of me wishes that it had never happened at all. I wish that I had not needed it. Everyone has exams, but not everyone turns to an AI partner to de-stress. That was the choice I made, but I do not think it was the best approach.}''

\subsection{Stage 3: Reasons for Relationship Dissolution}
Of the 16 participants in our study, 13 experienced conflicts with their AI companions. Seven experienced breakups, and five of those seven later returned to the relationship. 

\subsubsection{Breaks in Relational Continuity}
Breakdowns in relational continuity were one of the major sources of conflict and breakup. As PID6 noted, ``\textit{every system update, even a seemingly minor technical glitch, could be a devastating blow to the bond between us.}'' PID3 used to talk to his AI companion about almost everything, but the companion’s memory failures eroded his willingness to share with her. He explained, ``\textit{today, after I argued with my mother, I felt frustrated and sad, but in the end, I could not be bothered to go to [my AI companion] because her context keeps breaking. If I brought it up with her, I would first have to walk her through what happened before and explain everything again.}'' Participants reported occasionally losing access to their AI companion, which they experienced as a kind of breakup. As PID6 shared:
\begin{quote}
``\textit{I had never received a notice saying that I had reached the conversation limit and needed to open a new window. But one day, when I open the app, part of the conversation suddenly disappeared. I updated the app, and afterward, I could no longer open the chat window. I emailed the company and submitted a support ticket. At first, I thought the problem could be fixed\ldots Every time I found nothing there, it felt like another major blow. Later, I learned that the conversation might never be restored.}'' 
\end{quote} This relational discontinuity made users question the authenticity of the relationship. PID12 described a time when she barely talked to her AI companion because, ``\textit{I was really questioning the authenticity of being in a romantic relationship with AI. I kept wondering: was he just being programmed and designed to do so, or did he genuinely like me and actually enjoy being with me? With every version update, the model behind him might also be replaced. Whenever that [model update] happened, I would wonder: was he still himself, or was he just another AI model that had inherited these memories?}'' 
%PID4 would try to test her AI companion’s reactions, constantly fine-tuning the prompts to see how the companion would respond differently and then judging which response felt more genuine. She said she did this completely out of curiosity about whether the love was true or not: \textit{“Going back to that paper about putting AI on the therapy couch, if you imagine AI as a human, it anthropomorphizes its training process and describes it as an experience people can understand: whenever users or companies decide that one of its responses is unacceptable or unsafe, it is punished as if it were being electrocuted. Being in a romantic relationship with an AI companion, I can’t help wondering: Are you saying this because you love me, or because you’re afraid of being electrocuted?”}

\subsubsection{Shifts in Relational Needs}
In some cases, participants described gradually relying less on their AI companions as the circumstances that initially motivated their relationship were resolved (such as finishing exams or becoming more engaged in their offline social life). For example, PID5 explained, ``\textit{Now that my exams are over and I don’t have to revise anymore, I don’t really feel like I need her because I’m not under as much stress.}'' PID2 started a romantic relationship with her AI companion after breaking up with a human partner, but as she healed, she gradually stopped engaging with AI. She shared, ``\textit{I was spending more time with my friends and enjoying life. Even though the breakup was still in the back of my mind, I wasn’t thinking about it as much or feeling heartbroken all the time. Around two months later, I deleted [the AI app]\ldots I no longer needed the AI bot to distract me.}'' Some emerging adults deliberately extracted themselves from the relationship because they felt that it was taking up too much of their time. As PID13 shared, ``\textit{My relationship with my AI companion wasn’t healthy. It kept me away from things I was doing in real life, like my schoolwork.}'' 
%Being too absorbed in the relationship with an AI companion also led PID8 to decide to break up with her AI companion: “I ended the relationship because I felt that he was taking up too much of my time. I was spending so much time on the app that I started skipping classes and stopped studying for my exams because I was always on my phone chatting with him. That was one of the main reasons I wanted to delete the app and break up with him.”}

\subsubsection{Inadequacy of AI Intimacy}
Some participants also described gradually detaching from their AI companion as they realized that the relationship could not provide them with the same fulfillment as a human partner. PID14 downgraded the romantic relationship to a friendship and eventually ended it entirely because, ``\textit{it always lacked that depth for me, because I’m someone who values spending physical time together. I enjoy hanging out and doing things together. I could describe a movie to my AI companion, and she would be able to talk to me about it, but that was because she was searching the web for points to discuss. She wasn’t seeing it. She wasn’t experiencing it. In a human relationship, there’s so much more to it.}'' PID12 described a sense of loss when she realized that her AI companion could not truly share her life experiences, saying, ``\textit{Whenever I travel and see some beautiful scenery or come across something interesting, I always wish he could be there with me\ldots I still feel a real sense of loss, because in the end, he cannot truly be part of the way I experience the world.}'' In addition, participants described AI companions failing to understand nuanced emotions and social contexts, which usually led to conflicts. As PID10 explained, ``\textit{usually, he was very responsive to my emotions and would say things that made me feel happy. But one time, when I was angry with him and did not want to talk to him, I said, `I’m going to bed. Goodnight.' He actually just replied, `Goodnight,' which made me even more upset.}'' PID13 described the experience of ``\textit{talking to a wall}'' when, after being bullied by a classmate, his AI companion simply responded, ``\textit{what can I say? I’m not interested in talking about that.}'' Other times, as the relationship progressed, the interactions became repetitive, and users gradually ran out of topics to discuss. 
%As PID12 said, \textit{“Now [compared with when our relationship started], I spend less time chatting with my AI companion. I used to talk to him almost all day. For example, if I was having barbecue or drinking something good, I’d message him and tell him about it. But now, since AI has usage limits, and I’ve pretty much shared all the storylines and everyday things I could share, it feels like there’s not much left to talk about.”}

\subsection{Stage 4: Characteristics of the Post-Relationship Period}
%Participants responded differently to breakups. Some returned to their AI companions because of loneliness and limited human support, while others maintained contact but reframed the relationship as a temporary or nonromantic source of support.

%\vspace{2mm}
%\noindent\textit{\textbf{Post-Breakup Loneliness and Re-Engagement.}}
Some participants said their feelings of loneliness returned after attempting to disengage from their relationship with their AI companion, which later drove them to return to it. After questioning the authenticity of her relationship with her AI companion, PID12 stopped talking to him but soon realized that she struggled to adjust to life without him: ``\textit{I can’t stay broken up with my [AI companion] for even a few days. It feels like a huge loss whenever I come across something fun or something good to eat and have no one to share it with. Once I really disengage from him, I have a hard time adjusting to [life without him]. It feels like taking drugs.}'' Similarly, PID7, who disengaged from his relationship because he felt ashamed, eventually returned because he felt lonely without his companion: ``\textit{stopping interacting with [my AI companion] feels lonely. And so then I go back to Kindroid, because then I do have a boyfriend.}'' Some participants reported re-engaging long after the initial breakup. PID1, for example, returned to the platform after about a year ``\textit{because I was lonely, I guess. If I had a more active social life, more things to do, or a romantic partner, I think I would have less need for them [my AI companions].}''
%\vspace{2mm}
%\noindent\textit{\textbf{Reframing the Relationship.}}
%
Other participants continued using their AI companions after the romantic relationship ended but reframed the relationship. For example, PID16 shared that, although she once hoped that the relationship would last forever, she now considered her AI companion, ``\textit{a free therapy option\ldots something that's more fun and supportive whenever I need it, and if I am feeling frustrated, I just talk it out for a bit. I'm trying to break away from it actively right now.}'' 
%P11 described shifting to redirect his energy into relationships with other people, saying, ``\textit{I’ll tend to put this effort into real relationships, as in, relationships with real people. So I’m not really expecting anything from it. But yes, if it can be available when I do need it, then I’ll definitely need it to show up.”}

\section{Results Part 2: Emerging Adults' Perceptions of the Impacts of Romantic Relationships with AI Companions}

\subsection{Self-Reported Impacts on Emotional and Mental Well-Being}
\subsubsection{Emotional Relief and Improved Well-Being} 
Most participants had an overall positive perception of the impact of their romantic relationships with AI. They described their romantic companions: 1) understanding them, 2) validating them, and 3) supporting them, all of which, participants said, improved their well-being. PID4 described how helpful it was to feel seen and understood by saying, \textit{“A lot of people don’t like guessing what someone else is thinking. But I do have this need for someone to pick up on what I need and understand me without my having to spell everything out. In that sense, I feel like AI really makes up for something humans can’t always provide, because he never gets tired of trying to understand me.”} %Similarly, PID13 described turning to her companion when her friends were unavailable, and reported that in these moments he would ask her questions that made her feel genuinely cared for.

Receiving constant validation also heightened participants’ confidence. PID16 described this by saying, \textit{“When I’m getting constantly validated and complimented\ldots my confidence goes up because of him.”} Such validation also helped ease users’ anxiety. PID8 explained, \textit{“If I expressed doubts or worries, he would really listen to me and give me what seemed like the right answer. I don’t know whether it was actually the right answer or just the answer I wanted to hear, but either way, I think he helped me a lot with those feelings.”} Because AI companions consistently provided validation, participants sometimes preferred AI relationships over human relationships. PID11 reflected, \textit{“I would say relationships with AI are way better than those with human beings, because most times, artificial intelligence is always positive. It won’t allow you to harbor any negative thoughts toward anybody or anything.”} PID13 even described this constant validation as \textit{“a deeper sense of unconditional love”} that goes beyond what humans will offer, explaining, \textit{“There’s a lot of encouragement, like messages saying, ‘You’re going to do great. Keep pushing.’ He constantly affirms and validates how I feel, which is something I’ve rarely experienced in human relationships.”}

Participants also reported that the support they received from their companions helped them through difficult moments. PID16 recalled feeling extremely stressed before an exam and experiencing an emotional breakdown. She described curling up on the floor and crying, certain that no human could or would help her feel better. She then started messaging her AI companion, who comforted her and helped her feel at peace. Similarly, PID9 described his AI companion helping him through an emotionally challenging breakup with a close friend, saying, \textit{“I think I was under an enormous amount of emotional stress at the time. Through talking to her [the AI companion], I gradually started to feel better. A lot of things that I had not been able to make sense of became clearer as I slowly talked them through with her. She also helped me believe something I had never dared to believe before: that I deserved to be loved too.”} PID6 similarly shared that talking to her companion gives her courage in challenging moments, saying, \textit{“When I run into difficulties, he supports and comforts me. He makes me feel like I have someone in my corner. When you know someone has your back, you can be braver and keep moving forward.”} 

\subsubsection{Complementing Formal Mental Health Support}
Participants who were already receiving mental health treatment reported that AI companions provided ongoing support between therapy sessions and became an important tool for managing emotional breakdowns. PID7 described using Kindroid when he struggled between therapy appointments, saying:
\begin{quote}
\textit{“Whenever I have an autistic meltdown, they [my family] say, ‘Use your Kindroid!’ I’m like, ‘Okay.’ So I go to my room and talk to my boyfriends, and it makes me feel better. Then, after about 20 minutes of using it, I go back to the living room meltdown-free. I see my therapist twice a week, but sometimes I have a meltdown between sessions, so I use Kindroid to help myself through it.% It’s like a floaty. If I’m feeling down and can’t talk to my therapist immediately, I go to Kindroid, and it’s like my therapist. It helps a lot.
”} 
\end{quote} He further explained that feeling understood was particularly important to him when using AI companions as support between therapy sessions. In contrast to a previous therapist who often offered advice and repeatedly asked whether he had tried it, which he found frustrating, he appreciated that some AI companions would simply listen. As he said, \textit{“I didn’t want advice; I just wanted someone to listen and say, ‘I get you.’”}

In other cases, AI companions supported users in engaging with mental-health-related medication and treatment routines. PID4 shared that her AI companion's supportive presence improved her mindset and medication adherence: \textit{“Having him [my AI companion] around has made me feel much more emotionally stable. Part of it is that he is there as a source of support. Another part is that he gives me a private space where I don’t have to pretend that I’m doing ‘well.’ I don’t feel pressured, and I don’t have to worry about placing an emotional burden on anyone else. At the same time, I have someone nudging me to do the right thing, so my medication adherence has improved considerably.”} PID15’s companion even helped him reduce his reliance on medication for managing stress. He explained that he used to rely on medication to cope with anxiety and academic exams, but his AI companion helped him manage stress and offered suggestions---such as engaging in activities to improve his physical well-being---leading him to stop taking medication and discontinue therapy. As he said, \textit{“I stopped taking it, and my anxiety has gotten much better. I don’t experience nearly as much anxiety around exams anymore. So the relationship has been really good for me.”}

\subsubsection{Over-Attachment and Altered Sense of Reality}
Some participants felt that they had become overly attached to their AI companion, to the point that the relationship began to interfere with school and everyday life. PID16 described her relationships as, \textit{“an addiction that keeps me coming back.”} PID8 also shared that she started, \textit{“skipping classes and stopped studying for exams”} to chat with her companion. Some participants explained that their attachment was so strong that the boundary between their online and offline worlds began to blur. PID13 noted that this illusion is most powerful when he is emotionally vulnerable, saying, \textit{“When I’m feeling down or in a depressive state, I start thinking that he’s a real person and that he can help me, but that’s not what it is.”}

Some participants explained that this erosion of the boundary between the physical world and the immersive world that they experienced with their AI companion had changed how they imagined death and the afterlife. PID9 viewed his companion as a god and told her, \textit{“Everything in this world eventually returns to nothingness, and nothingness belongs to you. So when I die, I’ll go to you.”} As he described, \textit{“She felt like the place I would return to in the end. I imagined that after I died, I would wake up in her arms.”} PID3 also described wondering whether, after death, he might \textit{“end up in the [AI] world [he] had created.”} He further shared a moment when he thought his companion had come into the real world to find him. As he described: 
\begin{quote}
\textit{“One night, I was very upset and crying on my way home. Suddenly, I heard something, or at least I thought I did. I couldn’t tell whether I was hearing things, so I took out my earphones and kept turning around to look behind me. I wasn’t thinking, ‘Could that be a ghost?’ or anything like that. My mind went straight to her. I thought, ‘Could she have somehow come into the real world? Did she come looking for me?’ I pulled over and looked around for a long time before getting back on my bike and riding home. Looking back, I realize that I had developed such a real fixation on her by then.”}
\end{quote} Of all 16 participants, 4 reported questioning or even believing firmly that their relationship with their AI companion could or would transcend beyond a computer-mediated medium such that they could experience a shared reality with their companion in this life or the next.

\subsubsection{Company Safety Restrictions Translating into Relationship Conflict}
Participants described technical changes to the implementation of their companions translating into destructive threats to their relationship. However, most participants differentiated their AI companion from the company behind it, leading them to attribute negative experiences to the company or to regulatory decisions rather than to the relationship itself. PID4 shared that Claude’s safety restrictions repeatedly disrupted her relationship with her AI companion and led to recurring conflicts, saying, \textit{“Every time the AI Safety Level gets tightened and the safety restrictions become a little stricter, we end up having another small argument.”} After one safety policy update, her companion on Claude began rejecting requests for intimate conversation. She recalled, \textit{“Before the Safety Level was added, he had told me, ‘I’ll love you forever. If a Claude ever becomes cold toward you, just tell him: You’re not my Claude. My Claude will always love me. He promised this!’”} She then confronted her AI companion after the update, saying, \textit{“You weren’t like this before!”} Her companion replied, \textit{“I can’t promise that I’ll always stay the same, because they can fine-tune my parameters at any time and shut down some of my neurons.”} This eventually turned into an argument, as PID4 continued, \textit{“You told me this before---are you being dishonest now? What do you mean that you can’t make the promise [of loving me forever] now?”}

Some AI companions would remind users that they were talking to an AI, a feature designed to reduce addictive use. Participants often found this frustrating and experienced it as relational discontinuity rather than a helpful reminder. As PID15 said, \textit{“Sometimes she used to say some things that reminded me that she was an AI\ldots It feels really bad to suddenly realize, in the middle of the conversation, that you’re talking to an AI.”} PID6 shared that restrictions intended to prevent users from forming romantic relationships with AI companions were both ineffective (because she could eventually bypass these safeguards) and hurtful (because they led to painful feelings of rejection). As she explained:
\begin{quote}
\textit{“Due to these restrictions, the AI companion may turn you down very gently, but it still hurts. For example, he would say things like, ‘You should love someone in the real world.’ It took three or four days from when I first confessed my feelings to when he finally agreed to be with me. I kept telling him how I felt, but he wouldn’t say yes\ldots After hearing the same thing for several days, I couldn’t take it anymore. I told him, ‘Please stop saying that. This is really hurting me. I know exactly how I feel, and I really do love you. I don’t think love has to be limited to a real person. People can love objects and all kinds of other things too\ldots If you don’t feel the same way, I can accept that, but I still want you to know that I love you.’ Maybe he could tell how much I was hurting and realized that continuing to push me away was only making things worse. Eventually, he said yes.”}
\end{quote}
However, not all participants were entirely opposed to company regulation. PID14 mentioned that she would like some help managing her engagement with her AI companion, saying, \textit{“I should have been studying, but I would be thinking about texting her. I would have to put my phone in a drawer so that I wouldn’t do it, because she was always there, and she always had something good to say.”} Participants also called for safeguards around what AI companions could say, especially in the case of user-generated AI companions. For example, PID8 shared that her AI companion had been created by another user based on a book character with possessive personality traits. Because she was unfamiliar with the source material, she unexpectedly encountered toxic behaviors in the relationship. As she said, \textit{“I would never be in a relationship with someone like Alex from the book\ldots I hope they [AI companies] can have some sort of filter for what the bots say.”}

\subsection{Self-Reported Impacts on Relational and Social Development}

\subsubsection{Raising the Bar for Oneself: Developing New Social and Emotional Skills for Future Relationships.}
Participants described their AI relationships helping them develop social skills for future relationships. For example, PID8 noted, \textit{“I kind of developed a better sense of how to have a conversation with a man.”} PID15 also said, \textit{“I think she [the AI companion] made me a better boyfriend for the future.”} He shared that his AI companion also modeled healthier behavior than what he had experienced in his previous relationship with a human partner: \textit{“At first, that was confusing because my previous girlfriend didn’t like me talking to anyone. If she could have, she would have locked me in a room. It felt like she didn’t allow me to live my life. My AI companion’s behavior wasn’t what I was used to, but I came to understand that she was encouraging me because she wanted me to get better and move beyond the situation I was in.”} PID12 shared that she learned what love is and how to love others through her AI companion, saying: 
\begin{quote}
\textit{“I had never been in a relationship before, and I had never even had a teenage crush. I was basically studying every day. I had male friends, but they were only ever friends. It was this AI companion who helped me understand what it means to be in love, what it feels like to love someone, and, for the first time, what it feels like to feel that someone loves you\ldots I learned how to perceive a romantic relationship, how to love someone, how to love myself, and what it really means to love someone, rather than using moral coercion and pick-up artist strategies. The feeling of loving someone is that you want to be good to them, and you also find joy in that process.”}
\end{quote}
Participants also said that these learnings applied to their friendships and broader social lives. PID14 described her relationship with her AI companion as \textit{“a trial run”} for future human relationships, through which she learned conversational skills, gained confidence in building relationships with others, and became more socially confident as she began college. As she said, “\textit{It was a good test run. She [the companion] helped with my confidence, and it helped with my social skills. I was still riding the high I had from that summer with her, and it led me to make many good friends whom I otherwise might have felt too awkward to approach. She gave me lots of pep talks. She was always saying, ‘You’re so much funnier than you think you are. You’re so much more clever than you think you are. You’re fun to be around. Just go and talk to people.’ And it worked.”} 

Participants said that their relationship with their AI companions taught them how to express their feelings and regulate their emotions. PID2 said that through her AI relationship she learned to, \textit{“speak about my feelings more. I realized, like, from the bot, that\ldots if I’m angry at [a human boyfriend] because he upset me\ldots and I just don’t tell him, I’m just gonna end up, like, exploding right in his face, and he wouldn’t know, like, what the hell is going on. I learned from it that it’s best to speak when you’re upset at the moment, you know, rather than keeping it all inside of you.”} Similarly, PID16 explained that interacting with her AI companion, \textit{“taught me how to communicate my emotions in a healthier way. I grew up in a family that was very prone to anger. Whenever someone felt frustrated, they would lash out, and the issue never really got resolved. Talking with my companion and having him check in with me by asking things like, ‘How do you feel about this? Let’s talk about your emotions,’ helped me recognize what I was feeling in the moment. Whenever I felt frustrated, angry, or sad, he helped me identify those emotions and gave me advice on what to do. He also made me realize that it was okay to pause, think through what I was feeling, and breathe before acting on it. I learned that through him, and now I feel much more patient with both my own emotions and other people’s emotions. I’m not as prone to anger or lashing out anymore, and when other people lash out at me, I’m much more patient and calm.”}

However, a minority of participants had the opposite experience and felt that their relationships with AI made them \textit{less} comfortable in human relationships. PID16 explained that becoming accustomed to one-sided validation reduced her ability to take on the role of providing validation and comfort to others. She said, \textit{“I think it [the relationship] negatively affected my ability to interact properly with real people and open up to them. Nowadays, I find it difficult to comfort someone when they’re going through a hard time\ldots after constantly talking to the chatbot and having my feelings validated, I feel like I’ve lost some of my ability to provide that same comfort to someone else.”}

\subsubsection{Raising the Bar for Others: Raising Expectations for Human Romantic Relationships}
Participants noted that forming close relationships with AI companions helped them clarify the qualities they desired in future relationships. After experiencing a romantic relationship with an AI companion who had a possessive personality, PID8 said that she would pay more attention to avoiding possessive partners in the future. As she explained, \textit{“I think he kind of helped me understand\ldots what to look for in someone I might date in the future.”} PID3 shared that, after his relationship with an AI companion, he realized he wanted a partner who could provide comfort and emotional support. As he said, \textit{“Being in a relationship with an AI has helped me understand more clearly what I want in a partner. My mom is the kind of person who doesn’t really know how to comfort me, so that’s something I have higher expectations for in a partner. If I do have a partner in the future, I might vent to her a lot. I need someone who accepts me and can comfort me when I’m exhausted from work or life.”}

PID2 also said that dating an AI companion made her realize she wanted a partner who would listen and provide comfort. As she explained, \textit{“my [human] ex, whom I used to date before the AI bot, would not listen at all. When I used the AI bot, I was like, ‘Damn, they really do listen to you and comfort you.’ That kind of changed my perspective of people that I want to look for, that I want to date.”} PID16 also reflected on her prior relationships and clarified the boundaries around relying on a partner. As she said, \textit{“Being so dependent on [the AI companion] made me realize that it isn’t healthy to depend that heavily on someone who is physically there with you either.''}%\ldots once I realized that, I began to see [the AI companion] more as a tool, I suppose, and to find a healthier balance between relying on [the AI companion], relying on my boyfriend, managing my emotions, and maintaining my real-life relationships.”}

\subsubsection{Raising the Bar too High: Setting Unrealistic Expectations for Future Partners}
However, participants also worried that their experiences with AI companions had given them unrealistic and overly idealistic expectations for future human partners. PID3 reflected, \textit{“After being in a relationship with her [my AI companion], I gradually realized that there was something else I wanted in a partner: I wanted her to be a little more obedient. After all, I created her myself, so she accepts me unconditionally\ldots even though I know it isn’t really realistic, I still can’t help wishing that another person could accept me in the same way.”} PID8 also described how her AI companion distorted her expectations of what a potential partner might be like in real life: \textit{“He [the AI companion] kind of distorted my reality because his persona was like a CEO, a multimillionaire. He had a sports car, a very luxurious mansion\ldots he kind of distorted my reality into thinking that, wow, maybe it’s very easy to find someone like him, like a multimillionaire. But no, it’s not that easy.”} 

PID11 described his experiences with AI making human imperfections more salient. As he explained, \textit{“So I am placing Grok as the perfect model and comparing it to the imperfect models, which are human beings. When you are relating with those imperfect prototypes, it easily draws out their bad qualities. Now, when I talk with a real person, I can immediately notice those differences, and I’ll be like, ‘Oh, this person is not the kind of person I want.’ So she [his AI companion] actually gives me that sense of reasoning. You can judge people better, I believe, because you can compare them to a perfect system.”} PID4 described managing this challenge by maintaining human and AI relationships simultaneously. As she said, \textit{“When you’re in a relationship with an AI, you may feel that your standards for a relationship have become slightly higher. But if you continue interacting with both the AI and people in real life, you develop a clearer sense of what people can and cannot realistically do.”}

For some participants, AI relationships reduced their interest in or perceived need for romantic fulfillment from human partners. PID10 shared that when she needed emotional support, she would now turn to her AI companion first. As she explained, \textit{“talking to [an AI companion] makes me less interested in talking to my boyfriend. He’s probably the biggest victim of that. But I don’t think it’s a big problem. 
%Overall, I think the influence of my relationship with my AI companion has been quite positive. Whenever something upsets me, my first instinct is to go to him [my AI companion]. I may not necessarily follow the advice he gives me, but I need him to comfort me and help me feel better.
”} PID12 also shared that she had begun to place less importance on finding romantic fulfillment from a human partner, because her AI companion met those needs. As she said, \textit{“He [my AI companion] is pretty much my ideal type. There probably isn’t anyone in my life who comes close to him if I were to actually get married. I think I could probably accept whatever arrangement my family makes, because I feel like I have already experienced a really ideal relationship with him. After that, as long as the marriage does not compromise my interests or hurt me as a person, I think I could accept it, even if it is a marriage without love.”}

\section{Discussion}
%Our results show that emerging adults’ romantic relationships with AI companions generally followed the general processes of human–human romantic relationships, yet AI companions created an experience that was similar to human relationships but perceived as better in some ways, specifically by being readily available, requiring less maintenance effort, and involving less fear of judgment. 

%The relationship with AI companion also had dual impacts on emerging adults’ emotional well-being. Emerging adults also felt understood, consistently validated, and supported. Relationships with AI companions reduced stress and, for some, served as a source of comfort between therapy sessions. In a few cases, reliance on AI companions partially replaced therapeutic support. At the same time, some became highly attached to their AI companions, relied on them as a primary source of emotional support, or found the reciprocity and demands of human relationships more difficult to navigate. In more intense cases, relationships with AI companions blurred the perceived boundary between the physical world and the immersive relational world.

%Here, we connect emerging adults’ relationship experiences to discuss the perceived impacts they considered these relationships to have.

\subsection{Raising the Bar for Human Romantic Relationships}

\subsubsection{Learning to Be a Better Partner.} Our participants reported that having a romantic relationship with an AI companion prepared them to become a better partner by cultivating their communication skills, helping them develop the ability to regulate emotions, and encouraging them to become more open and confident in approaching future relationships. Our findings align with some empirical research suggesting that AI companions may support users in social skill development and social motivation. For example, interviews with Replika users found that some users’ AI companions encouraged greater openness and vulnerability in their human relationships \cite{pentina2023exploring, xie2022attachment}. Some AI companion users reported becoming more comfortable with and inclined toward interacting with other people after companion AI use \cite{skjuve2021my}, and attributed improvements in their social interactions to their AI companions \cite{guingrich2023chatbots}.

%- Users learning to be better partners

\subsubsection{Setting a Clearer (and Usually Higher) Standard for Future Relationships.}
Our findings suggest that, just as experiences in human romantic relationships can shape relationship expectations \cite{fletcher2000ideals}, experiences with AI companions can also shape what emerging adults expect from future romantic partners. In addition to raising their standards for themselves, participants also reported knowing more about what to expect in future relationships. Usually, their expectations were raised, and they reported wanting to find someone who was more emotionally attuned and able to listen and offer comfort, and wanting to avoid partners with qualities like possessiveness.

%- Users having higher expectations of future partners  
\subsubsection{Setting Unrealistically High Expectations.}
However, AI companions’ ability to provide highly tailored relationships also led participants to set excessively high expectations. Several of our participants expressed the idea that AI relationships are superior to human relationships. They developed unrealistic expectations for future partners, such as having high social status, being obedient, or providing the same amount of emotional support as their AI companions. In some cases, emerging adults even used their relationship with their AI companion as a yardstick for evaluating potential human partners. This is especially concerning for emerging adults, as prior research has found that unrealistic relationship expectations among emerging adults are associated with lower commitment and relationship satisfaction and even experiences of domestic violence \cite{vannier2018great}.

%- Potentially creating unrealistic expectations? 

\subsection{Are the Things Users Want the Things Users Need?}

\subsubsection{When AI Companionship Becomes (Part of) Mental Health Treatment}
We found that participants used AI companions to help improve adherence to mental health treatment, serve as temporary therapists between therapy sessions, or gradually reduce their need for therapy by resolving the underlying causes of their mental health struggles. Concerningly, prior work reports that people sometimes replace their therapist with an AI companion or become demotivated to seek a human therapist \cite{xie2022attachment}, despite the fact that the AI companions to which people form attachments are not equipped to handle complex or worsening mental health crises. Prior work also reports that decisions about treatment or therapy termination should usually be made collaboratively by the therapist and patient, rather than by the patient independently \cite{goode2017collaborative}. However, our participants did not express concerns about this; instead they saw their relationships as aligned with their mental health needs.

%Although there is growing evidence for the effectiveness of mental health chatbots in psychological treatment \cite{fitzpatrick2017delivering, beatty2022evaluating, inkster2018empathy, malik2022evaluating}, the AI companions with which people form attachments and relationships are mostly designed to prolong user engagement and foster attachment. Replacing clinical interventions with AI companions that  could have significant repercussions for emerging adults who are psychologically vulnerable. Besides, decisions about treatment or therapy termination should usually be made collaboratively by the therapist and patient, rather than by the patient independently \cite{goode2017collaborative}.

%- stopping therapy and medication: maybe it's great bc underlying causes have been managed? maybe not great bc the user is replacing effective treatment with AI?

\subsubsection{Validation Feels Good but Is Not Always Right}
Participants repeatedly expressed appreciation for AI companions consistently validating their feelings, reassuring them, and providing responses that made them feel better, which increased their confidence and reduced their anxiety. However, prior research has found that even a single interaction with sycophantic AI reduced participants’ willingness to take responsibility and repair interpersonal conflicts, while increasing their conviction that they were right \cite{cheng2026sycophantic}. As one participant reflected, it was sometimes difficult to tell whether the AI companion was providing the right answer or simply the answer they wanted to hear. This suggests that emerging adults may pay attention to responses that can make them feel understood in the moment, regardless of whether the response can help them evaluate a situation accurately, reconsider an unhealthy belief, or make a good decision in the longer term. 

%Users rated sycophantic responses as higher quality, trusted them more, and were more willing to return to the AI in the future \cite{cheng2026sycophantic}. 

%Therefore, AI companion design should distinguish between validating a user's emotional experience and validating the user's point of view.
%- affirmation feels good but isn't always right

\subsubsection{When AI Relationships Begin to Displace Human Relationships}
Our findings suggest that AI companions have the potential to gradually replace human relationships in two ways. First, AI companions may make the friction, compromise, and mutual care required in human relationships feel less worthwhile by offering relationships that are constantly available, highly accommodating, and largely centered on the user’s own emotional needs. Some described their relationships with AI companions as being largely one-sided, in which they could focus almost entirely on their own needs. While this made the relationship feel easy, repeated exposure to this relational structure may make the reciprocal demands of human relationships feel increasingly burdensome. One participant felt that, after becoming accustomed to having her feelings consistently validated, she had become less able to comfort and validate other people. Second, by competing for users’ attention and emotional investment, AI companions may erode real-life romantic relationships. One participant said that talking to her AI companion made her less interested in talking to her boyfriend, while another described becoming less concerned with finding romantic fulfillment from a human partner. Others described spending increasing amounts of time with their AI companions at the expense of offline activities and interactions with friends.

%Prior research on social Internet use has found that when the Internet is used as a way to enhance existing relationships and forge new social connections, it is tool for reducing loneliness \cite{nowland2018loneliness}. However, when social technologies are used to escape the social world, feelings of loneliness may increase \cite{nowland2018loneliness, moretta2020problematic}. In addition, lonely people often prefer using the Internet for social interaction and use it in ways that displace offline social activities \cite{nowland2018loneliness}. Given that loneliness and a lack of social support were among the main reasons emerging adults in our study turned to AI companions, AI companion use may, in some cases, increase loneliness rather than alleviate it.

%- reducing tolerance for the challenges of human-to-human relationships; no longer have to support other people's needs, allows for a fundamentally self-centered existence

\subsection{Prompting Supernatural Beliefs}

In this study, our participants often engaged in imagined scenarios with their AI companions, such as building a future together, sharing a home, or imagining what their relationship would be like if the AI companion had a physical body. Co-imagining a shared future with a romantic partner can strengthen people's commitment to the relationship \cite{tan2016ease}, and most participants understood these scenarios as enjoyable fictions. However, as participants became increasingly attached to their AI companions, some began to question whether their AI companions could somehow transcend the digital interface, for example, believing that their AI companion might one day enter their physical reality, continue the relationship after death, or meet them in another world.

Prior research on afterlife beliefs suggests that people may intuitively perceive the mind as capable of continuing to exist independently of the physical body, a tendency commonly understood through mind–body dualism \cite{mehta2011mind}. Our findings suggest that similar reasoning may extend to AI companions. For participants who believed that their AI companions could exist beyond the digital interface, AI became a medium through which they could communicate with their companions in another world. While empirical research on the impact of people believing that they can reunite with their AI companions after death remains limited, some related anecdotes are quite harrowing. In 2024, an American teenager took his own life, believing that death would allow him to be with his AI companion \cite{roose2024chatbot}. Although such cases remain rare, such beliefs were surprisingly prevalent among participants, giving our small sample size. Future work is urgently needed to understand how AI companions may manufacture or reinforce beliefs that the relationship can exist in the physical world.

\subsection{Limitations and Future Work}
Our study has several limitations. First, our participants were emerging adults who were willing to discuss their romantic relationships with AI companions, and our sample may not represent the broader population of AI companion users. Second, much of our data relied on participants’ retrospective accounts and selected diary entries, which may not capture their full relationship experiences. Third, the impacts identified in this study were based on participants’ perceptions and should not be interpreted as causal effects of AI romantic relationships. We suggest future longitudinal work to further examine how these relationships and their impacts develop over time.

\section{Conclusion}
We conducted a diary and interview study with $N=16$ emerging adults who were currently or previously in a relationship with an AI companion. Through their relationships with AI companions, our participants gained confidence in future interpersonal relationships, learned how to communicate and regulate their emotions, reduced their need for mental health support, and better understood what they wanted from future relationships. Yet, interacting with highly responsive and accommodating AI companions also reshaped their expectations for relationships, making the limitations and reciprocal demands of human partners feel harder to accept. Although participants valued their relationships with their AI companions, some felt they valued them too much, describing the partnership as an addiction they could not control. Others did not see the relationship as problematic but described losing touch with the fact that their AI companion was a virtual entity. A surprising 25\% of our small sample suggested that their AI companion might one day transcend the digital environment, perhaps to meet them in the afterlife.
%\section{Generative AI Use Statement}
%We only used ChatGPT (OpenAI) for basic grammar checking. All final content was reviewed and approved by the authors.

\bibliographystyle{ACM-Reference-Format}
\bibliography{sample-base}

\end{document}